\documentclass[preprint,pra,showpacs,superscriptaddress]{revtex4}
\usepackage{graphicx}
\usepackage{bm}
\usepackage{amsmath,amssymb}
\begin{document}

\title{Macroscopic entanglement of many-magnon states}

\author{Tomoyuki Morimae}
\email{morimae@ASone.c.u-tokyo.ac.jp}
\affiliation{Department of Basic Science, University of Tokyo, 3-8-1 Komaba, Tokyo 153-8902, Japan}
\affiliation{PRESTO, JST, 4-1-8 Honcho Kawaguchi, Saitama, Japan}
\author{Ayumu Sugita}
\email{sugita@a-phys.eng.osaka-cu.ac.jp}
\affiliation{Department of Applied Physics, 
Osaka City University, 3-3-138 Sugimoto, Osaka 558-8585, Japan}
\author{Akira Shimizu}
\email{shmz@ASone.c.u-tokyo.ac.jp}
\affiliation{Department of Basic Science, University of Tokyo, 3-8-1 Komaba, Tokyo 153-8902, Japan}
\affiliation{PRESTO, JST, 4-1-8 Honcho Kawaguchi, Saitama, Japan}
\date{\today}
            
\begin{abstract}
We study 
macroscopic entanglement of various pure states of
a one-dimensional $N$-spin system
with $N \gg 1$.
Here, a quantum state is said to be 
macroscopically entangled if it is 
a superposition of macroscopically distinct states.
To judge whether such superposition is hidden 
in a general state, we use an essentially unique index $p$:
A pure state is macroscopically entangled if $p=2$, 
whereas it may be entangled but not macroscopically if $p<2$.
This index is directly related to fundamental stabilities of 
many-body states.
We calculate the index $p$ for various states 
in which magnons are excited with various densities and wavenumbers.
We find macroscopically entangled states ($p=2$) as well as 
states with $p=1$.
The former states are unstable in the sense that 
they are unstable against some local measurements.
On the other hand, the latter states
are stable in the senses that 
they are stable against any local measurements
and that their decoherence rates never 
exceed $O(N)$ in any weak classical noises.
For comparison, 
we also calculate 
the von Neumann entropy $S_{N/2}(N)$
of a subsystem composed of $N/2$ spins as
a measure of bipartite entanglement. 
We find that $S_{N/2}(N)$ of some states with $p=1$ 
is of the same order of magnitude as 
the maximum value $N/2$.
On the other hand, 
$S_{N/2}(N)$ of the macroscopically entangled states with $p=2$
is as small as  $O(\log N)\ll N/2$.
Therefore, larger $S_{N/2}(N)$ does {\em not} 
mean more instability.
We also point out that these results are 
partly analogous to those for interacting many bosons.
Furthermore, the origin of the huge entanglement, 
as measured either by $p$ or $S_{N/2}(N)$, 
is discussed to be due to spatial propagation of magnons.
\end{abstract}
\pacs{03.67.Mn, 75.45.+j, 75.10.Jm, 03.65.Yz}
\maketitle  

\section{\label{sec:level1}introduction and summary}
Many-partite entanglement, i.e., 
entanglement in a system that is composed of many sites or parties, 
has been attracting much attention 
recently 
\cite{Bennett,SM,Ukena,Sugita,mermin,
Arnesen,Wang,Vidal,O'Connor,Subrahmanyam,Syljuasen,
J.Vidal,Verstraete,Osborne,Meyer,Stockton,Gunlycke,Sun,Zhou,Kamta,
Bose,Verstraete2}.
It is known that 
the number of possible measures of entanglement 
grows dramatically as the number of sites is increased \cite{Meyer}.
Different measures are related to different physical properties.
Therefore, one must specify physical properties of interest in order to
determine a proper measure or index.

In this paper, we study 
{\em macroscopic entanglement}
of various states in a quantum many-spin system.
Here, a quantum state is said to be 
macroscopically entangled if it is 
a superposition of macroscopically distinct states
(see Sec. \ref{macroentanglement}).
Although such superposition is trivially recognized for some states
(such as the `cat' state), 
it is hard to find such superposition by intuition for general 
states.
In order to judge whether such superposition is hidden 
in a general state, we use an essentially unique index $p$, defined by
Eq.\ (\ref{def-p}).
A pure state is macroscopically entangled if $p=2$, 
whereas it may be entangled but not macroscopically if $p<2$.
Unlike many other measures or indices of entanglement, 
there is an efficient method of computing $p$ for 
any given states \cite{Sugita}.

It was shown by Shimizu and Miyadera \cite{SM}
(hereafter refereed as SM) 
that this index is directly related to fundamental stabilities of 
many-body states, i.e., to fragility in noises or environments
and to stability against local measurements.
That is, a state with $p=1$ is not particularly unstable
against noises
in the sense that its decoherence rate does not 
exceed $O(N)$ in any noises or environments, 
whereas the decoherence rate of a state with $p=2$ can become 
as large as $O(N^2)$ \cite{O(N)}.
Furthermore, a quantum state with $p=2$ is unstable against local 
measurements, whereas a homogeneous state with $p=1$ is stable.

We consider a one-dimensional $N$-spin system with $N \gg 1$, 
and calculate the index $p$ for various pure states 
in which magnons are excited with various densities and wavenumbers.
We find macroscopically entangled states ($p=2$) as well as 
`normal' states with $p=1$ which 
{\em are} entangled but {\em not} macroscopically.
According to SM, they are unstable and stable many-body states, respectively.

For comparison, 
we also calculate 
the von Neumann entropy $S_{N/2}(N)$
of a subsystem composed of $N/2$ spins as
a measure of bipartite entanglement. 
We find that 
some states with $S_{N/2}(N)=O(N)$, which is of 
the same order of magnitude as 
the {\em  maximum} value $N/2$, 
are `normal' states in the sense that $p=1$.
On the other hand, 
some of other states, which are macroscopically entangled ($p=2$), 
have much smaller value of $S_{N/2}(N)$ of $O(\log N)$.

These results demonstrate that the degrees of entanglement are totally 
different if different measures or indices are used.
Furthermore, 
stabilities of quantum states are not simply related to 
the degrees of entanglement: Different stabilities are 
related to different measures or indices.
In particular, 
fragility in noises and
the stability against local measurements are directly related to $p$, 
hence are basically independent of $S_{N/2}(N)$.

The results also demonstrate that
states with huge entanglement, as measured 
by either $p$ or $S_{N/2}(N)$, 
can be easily constructed by simply applying creation operators of magnons
to a ferromagnetic state, which is a separable state.
Neither randomness nor elaborate tuning  
are necessary to construct states with huge entanglement
from a separable state.
This should be common to most quantum systems:
By exciting a macroscopic number of elementary excitations, 
one can easily construct states with huge entanglement.
To generate such states experimentally, however, 
one must also consider the fundamental stabilities mentioned above:
States with $p=2$ would be quite hard to generate experimentally, 
whereas states with large $S_{N/2}(N)$ would be able to 
be generated rather easily.

The present paper is organized as follows: 
In Sec.~\ref{magnon}, 
we shortly review physics of magnons, and 
present state vectors of many-magnon states under consideration.
We 
explain the index $p$ for the macroscopic entanglement, 
and present 
an efficient method of computing $p$ in Sec.~\ref{measure}. 
In Sec.~\ref{many-partite}, 
we study macroscopic entanglement of many-magnon states
by evaluating $p$.
We study 
their bipartite entanglement in Sec.~\ref{two-partite}
for comparison purposes.
Stabilities of the many-magnon states are discussed 
in Sec.~\ref{sec-stability}.
In Sec.~\ref{sec-BEC},
we point out that our results are 
analogous to those for interacting many bosons.
Furthermore, we discuss the origin of the huge entanglement
in Sec.~\ref{sec-origin}.

\section{Many-magnon states}\label{magnon}
In this section, we briefly review the physics 
of magnons in order to establish notations.

A magnon is an elementary excitation of magnetic materials. 
It is a quantum of a spin wave that is a collective motion 
of the order parameter, 
which is the magnetization $\vec{\mathcal {M}}$ for a ferromagnet. 

For example, consider a one-dimensional Heisenberg ferromagnet 
which consists of $N$ spin-$\frac{1}{2}$ atoms. 
Under the periodic boundary condition, its Hamiltonian is given by
\begin{eqnarray}
\hat{H}&=&-J\sum_{l=1}^N\hat{\vec{\sigma}}
(l)\cdot\hat{\vec{\sigma}}(l+1).\label{hamiltonian}
\end{eqnarray}
Here, $J$ is a positive constant, 
and $\hat{\vec{\sigma}}(l)\equiv(\hat{\sigma}_x(l),
\hat{\sigma}_y(l),\hat{\sigma}_z(l))$, 
where 
$\hat{\sigma}_x(l),\hat{\sigma}_y(l),
\hat{\sigma}_z(l)$ 
are Pauli operators on site $l$. 
We denote eigenvectors of $\hat{\sigma}_z$ corresponding to 
eigenvalues +1 and -1 by $|\uparrow\rangle$ and 
$|\downarrow\rangle$, respectively. 
One of the ground states of the Hamiltonian is 
$|\downarrow^{\otimes N}\rangle$, in which 
$\vec{\mathcal {M}}$ points to the $-z$ direction.
The state in which one magnon with wavenumber 
$k$ is excited on this ground state is 
\begin{eqnarray}
|\psi_{k;N}\rangle\equiv\frac{1}
{\sqrt{N}}\sum_{l=1}^Ne^{ikl}\hat{\sigma}_+(l)|\downarrow^{\otimes N}
\rangle,\label{onemag}
\end{eqnarray}
where 
$\hat{\sigma}_+(l) \equiv (\hat{\sigma}_x(l)+i\hat{\sigma}_y(l))/2$.
The excitation energy of $|\psi_{k;N}\rangle$ 
is easily calculated as 
\begin{eqnarray}
E_{k;N}=8J\sin^2\frac{k}{2}.
\end{eqnarray}
It goes to zero as $k\to0$ because a magnon 
is a Nambu-Goldstone boson. The dispersion 
relation for small $k$ is nonlinear because 
a magnon is a non-relativistic excitation. 
Because of the periodic boundary condition, 
$k$ takes discrete values in the first Brillouin zone, 
$-\pi < k \leq \pi$;
\begin{eqnarray}
k=\frac{2\pi}{N}j \quad(j:\mbox{integer},\
-\frac{N}{2}< j \le\frac{N}{2}). 
\end{eqnarray}
It is conventional to define the 
`creation operator' of a magnon with wavenumber $k$ by
\begin{eqnarray}
\hat{M}_k^{\dagger}\equiv\frac{1}{\sqrt{N}}
\sum_{l=1}^Ne^{ikl}\hat{\sigma}_+(l).
\label{Mdagger}\end{eqnarray}
The commutation relations are calculated as
\begin{eqnarray}
\left[\hat{M}_k^{\dagger},\hat{M}_{k'}^{\dagger}\right]
&=&\left[\hat{M}_k,\hat{M}_{k'}\right]= 0,\\
\left[\hat{M}_k,\hat{M}_{k'}^{\dagger}\right]
&=&-\frac{1}{N}\sum_{l=1}^Ne^{i(k'-k)l}\hat{\sigma}_z(l).
\label{commute}
\end{eqnarray}
When the number $m$ of magnons is much smaller than $N$, 
Eq.~(\ref{commute}) can be approximated as
\begin{eqnarray}
\left[\hat{M}_k,\hat{M}_{k'}^{\dagger}\right]
\simeq\frac{1}{N}\sum_{l=1}^Ne^{i(k'-k)l}=\delta_{k,k'}.
\label{boson}
\end{eqnarray}
Therefore, magnons behave as bosons when $m\ll N$. 

Using the creation operators, we define the $m$-magnon state 
with wavenumbers $k_1,k_2,\ldots,k_m$ by
\begin{eqnarray}
|\psi_{k_1,k_2,\ldots,k_m;N}\rangle
\equiv
\frac{G_{k_1,k_2,\ldots,k_m;N}}{\sqrt{n_a! n_b! \cdots}}
\prod_{i=1}^m\hat{M}_{k_i}^{\dagger}
|\downarrow^{\otimes N}\rangle.
\label{m-magnon state}
\end{eqnarray}
Here, $1/\sqrt{n_a! n_b! \cdots}$ is the usual normalization factor for 
bosons, where $n_\nu$ ($\nu = a, b, \cdots$) 
denotes the number of $k_i$'s having equal values, and
$G_{k_1,\ldots,k_m;N}$ is an extra normalization factor
which comes from the fact that magnons are not strictly bosons.
Without loss of generality, we henceforth assume that
\begin{equation}
k_1 \leq k_2 \leq \ldots \leq k_m.
\end{equation}
When $m\ll N$, the magnons behave as bosons so that 
$G_{k_1,\ldots,k_m;N} = 1$ and
\begin{equation}
\langle \psi_{k_1,k_2,\ldots,k_m;N} |
\psi_{k'_1,k'_2,\ldots,k'_m;N}\rangle
= 
\delta_{k_1,k'_1} \delta_{k_2,k'_2} \cdots \delta_{k_m,k'_m}
\end{equation}
to a good approximation.
On the other hand, the deviations from these relations
become significant when $m=O(N)$.

An $m$-magnon state
with a small density ($m/N\ll 1$) of magnons 
is an approximate energy eigenstate.
Although an $m$-magnon state
with a large number ($m=O(N)$) of magnons
is not generally a good approximation to energy eigenstate, 
such a state is frequently used in discussions 
on a macroscopic order
because many magnetic phase transitions can be regarded as
condensation of $O(N)$ magnons.
For example, the state in which 
$\vec{\mathcal{M}}$ points to a direction with the direction vector 
$(\sin\theta\cos\alpha,\sin\theta\sin\alpha,\cos\theta)$ 
can be described as 
\begin{eqnarray}
|(\theta\alpha)^{\otimes N}\rangle
&=&\left(e^{-i\alpha}\cos\frac{\theta}{2}|\uparrow\rangle+
\sin\frac{\theta}{2}|\downarrow\rangle\right)^{\otimes N}
\label{separable}\\
&=&\sum_{m=0}^Ne^{-im\alpha}\sqrt{B_m}|\psi_{(k=0)^m;N}\rangle,
\label{spon}\end{eqnarray}
where $|\psi_{(k=0)^m;N}\rangle$ is the 
$m$-magnon state with $k_1=\ldots=k_m=0$, 
and $B_m$ is the binomial coefficient;
\begin{eqnarray}
B_m&\equiv&
{N \choose m} \left(\cos^2\frac{\theta}{2}\right)^m
\left(\sin^2\frac{\theta}{2}\right)^{N-m}.
\end{eqnarray}
When $\theta \neq \pi$, 
$B_m$ has a peak at $m=N\cos^2\frac{\theta}{2} = O(N)$, and
thus a macroscopic number of magnons are 
`condensed'. 

Note that 
$|\downarrow^{\otimes N}\rangle$
and 
$|(\theta\alpha)^{\otimes N}\rangle$
belong to the {\em same} Hilbert space
because we assume that $N$ is large but {\em finite}, 
although they will 
belong to different Hilbert spaces if we let $N\to\infty$.
For the same reason, 
all $|\psi_{k_1,k_2,\ldots,k_m;N}\rangle$'s 
belong to the same Hilbert space even if $m = O(N)$.

\section{the index of macroscopic entanglement}\label{measure}
In this section, we present the index of macroscopic entanglement, 
and an efficient method of computing it. 
We also explain its physical meanings by giving a few examples. 
Relation between this index and stabilities of many-body states 
will be explained in 
Sec.\ \ref{sec-stability}.

\subsection{The index $p$}
\label{macroentanglement}
We are most interested in 
superposition of macroscopically distinct states,
which has been attracting much attention for many years
\cite{Leggett1,Leggett2,MQC1,MQC2,MQC3}.
We note that such superposition was defined rather ambiguously.
For example, the `disconnectivity' defined in Ref.\ \cite{Leggett1} 
is not invariant under changes of canonical variables, 
such as from the pairs of positions and momenta to the 
pair of a field and its canonical conjugate.
Furthermore, in much literature the energy scale is not specified
to determine the degrees of freedom involved in the superposition.
However, the degrees of freedom, hence the disconnectivity,  
usually grows (decreases) quickly with 
increasing (decreasing) the energy scale under consideration \cite{molecule}.
On the other hand, SM proposed a new definition that is 
free from these ambiguities.
We therefore follow SM.

We first fix the energy range under consideration.
For definiteness 
we assume that in that energy range the system can be regarded as 
$N$ spin-${1 \over 2}$ atoms, which are arranged 
on a one-dimensional lattice.
We note that two states are `macroscopically distinct' iff 
some macroscopic variable(s)  
takes distinct values for those states.
As a macroscopic variable, 
it is natural to consider 
the sum or average of
local observables over a macroscopic region \cite{current}.
Since the average can be directly obtained from the sum, 
we only consider the sum in the following.
That is, we consider additive operators \cite{TD}, which take the 
following form:
$\hat{A}=\sum_{l=1}^N\hat{a}(l)$.
Here, $\hat{a}(l)$ is a local operator on site $l$, 
which, for the spin system under consideration, is a linear combination 
of the Pauli operators 
$\hat{\sigma}_x(l),\hat{\sigma}_y(l), \hat{\sigma}_z(l)$ 
and the identity operator $\hat 1(l)$ on site $l$.
Since we will consider all possible additive operators, 
we do {\em not} assume that $\hat{a}(l')$ ($l' \neq l$) is 
a spatial translation of $\hat{a}(l)$.

Two states, $|\psi_1 \rangle$ and $|\psi_2 \rangle$,
are macroscopically distinct iff  
some additive observable(s) $\hat A$ 
takes `macroscopically distinct values' 
for those states in the sense that 
\begin{equation}
\langle \psi_1 | \hat A |\psi_1 \rangle
-
\langle \psi_2 | \hat A |\psi_2 \rangle
= 
O(N).
\end{equation}
Therefore, if a pure state $|\psi \rangle$ has fluctuation
of this order of magnitude, i.e., if
$\displaystyle
\delta A
\equiv
[\langle \psi | \Delta\hat{A}^\dagger \Delta\hat{A} |\psi \rangle ]^{1/2}
= O(N)
$
for some additive observable(s) $\hat A$, 
where $\Delta\hat{A} \equiv \hat{A} - \langle\psi|\hat{A}|\psi\rangle$,
then the state is a superposition of macroscopically distinct states.
On the other hand, 
if $\delta A = o(N)$ \cite{O(N)} for every additive observable $\hat A$
the state has `macroscopically definite values' for all  
additive observables.
A typical magnitude of $\delta A$ for such a state is
$\delta A = O(N^{1/2})$ \cite{Landau}.
To express these ideas in a simple form, 
we define an index $p$ for an arbitrary pure 
state $|\psi\rangle$ by the asymptotic behavior (for large $N$) 
of fluctuation of the additive observable that 
exhibits the largest fluctuation for that state \cite{pisthesame}:
\begin{equation}
\sup_{\hat{A}\in\mathcal{A}}
\langle\psi|
\Delta\hat{A}^\dagger \Delta\hat{A}
|\psi\rangle
=O(N^p).
\label{def-p}\end{equation}
Here, $\mathcal{A}$ is the set of all additive operators. 
According to the above argument, 
$|\psi\rangle$ is a superposition of macroscopically distinct states
iff $p=2$, and for pure states $p$ is the essentially unique index 
that characterizes such a superposition.
We therefore say that a pure state is {\it macroscopically entangled}
iff $p=2$.

\subsection{An efficient method of computing $p$}
It is well-known that many 
entanglement measures which are defined as an extremum 
are intractable 
\cite{Syljuasen,Verstraete,Osborne,Stockton}. 
In contrast, there is an efficient method of computing 
the index $p$ \cite{Sugita}.
We here explain the method briefly assuming an $N$ spin-$\frac{1}{2}$ system.

Any local operator $\hat a(l)$ 
on site $l$ can be expressed as a linear combination 
of $\hat{\sigma}_x(l), \hat{\sigma}_y(l), \hat{\sigma}_z(l)$ and $\hat{1}(l)$.
Since the identity operator $\hat 1$ does not have fluctuation for any state, 
we can limit ourselves to local operators that are
linear combinations of Pauli operators.
Therefore, an additive observable in question 
generally takes the following form;
\begin{eqnarray}
\hat{A}=\sum_{l=1}^N \hat a(l)
=
\sum_{l=1}^N\sum_{\alpha=x,y,z}
c_{\alpha l}\hat{\sigma}_{\alpha}(l),\label{aaa}
\end{eqnarray}
where $c_{\alpha l}$'s are complex coefficients (see Sec.~\ref{hermitian}).
Since the local operators should not depend on $N$
(because otherwise $\hat{A}$ would not become additive), 
$c_{\alpha l}$'s should not depend on $N$,
hence the sum 
$
\sum_l \sum_\alpha |c_{\alpha l}|^2
$
is $O(N)$.
Since we are interested only in the power $p$ of 
$\langle\psi| \Delta\hat{A}^\dagger \Delta\hat{A} |\psi\rangle = O(N^p)$, 
we can normalize $c_{\alpha l}$ without loss of generality as 
\begin{eqnarray}
\sum_{l=1}^N\sum_{\alpha=x,y,z} |c_{\alpha l}|^2
=N.\label{additive}
\end{eqnarray}
The fluctuation of $\hat{A}$ for a given state $|\psi\rangle$ is 
expressed as 
\begin{eqnarray}
\langle\psi| \Delta\hat{A}^\dagger \Delta\hat{A} |\psi\rangle
=\sum_{\alpha,l,\beta,l'}
c_{\alpha l}^* c_{\beta l'}
V_{\alpha l,\beta l'},
\label{flucbyVCM}\end{eqnarray}
where $V_{\alpha l,\beta l'}$ is the hermitian matrix defined by
\begin{eqnarray}
V_{\alpha l, \beta l'} \equiv 
\langle\psi|\Delta
\hat{\sigma}_{\alpha}(l)\Delta
\hat{\sigma}_{\beta}(l')|\psi\rangle,
\end{eqnarray}
which we call the {\em variance-covariance matrix} (VCM) for $|\psi\rangle$.
It is seen from Eq.~(\ref{flucbyVCM}) that 
eigenvalues of this matrix are non-negative,
and that 
$\langle\psi|\Delta\hat{A}^\dagger \Delta\hat{A}|\psi\rangle$
takes the maximum value when $c_{\alpha l}$ is an eigenvector of the VCM
corresponding to the maximum eigenvalue $e_{\rm max}$.
By taking $c_{\alpha l}$ of Eq.~(\ref{flucbyVCM}) as such an eigenvector, 
we obtain
\begin{eqnarray}
\sup_{\hat{A}\in\mathcal{A}}
\langle\psi|\Delta\hat{A}^\dagger \Delta\hat{A}|\psi\rangle
=e_{\rm max}N.
\end{eqnarray}
Therefore, $e_{\rm max}$ is related to $p$ as
\begin{equation}
e_{\rm max} = O(N^{p-1}).
\end{equation}
For example, 
$p=1$ if $e_{\rm max}=O(1)$ whereas $p=2$ if $e_{\rm max}=O(N)$. 

Note that we can evaluate $p$ of a given state using this method
in a polynomial time of the number $N$ of spins,
because we have only to calculate the maximum eigenvalue of a VCM, which 
is a $3N \times 3N$ matrix.

\subsection{Examples of macroscopically entangled states}
The $N$-spin GHZ state, or the `cat' state,  
$
|{\rm GHZ} \rangle
=
\frac{1}{\sqrt{2}}\left(|\downarrow^{\otimes N}\rangle+
|\uparrow^{\otimes N}\rangle\right), 
$
violates a generalized Bell's inequality by a macroscopic 
factor \cite{mermin}.
The index $p$ correctly indicates that this state 
is macroscopically entangled, $p=2$ \cite{Ukena}.
In contrast, $S_{N/2}(N)$ (which is defined by Eq.~(\ref{SN/2}) below)
of this state is extremely small; $S_{N/2}(N)=1$.
It may be intuitively trivial that this state is 
macroscopically entangled.
However, intuition is useless for more general states
such as the following examples.
The greatest advantage of using $p$ is that it 
correctly judges the presence or absence of 
macroscopic entanglement for any complicated pure states.

For example, 
it was recently shown \cite{Ukena} that 
the quantum state of many qubits in a quantum computer 
performing Shor's factoring algorithm
is transformed in such a way that 
$p$ is increased as the computation proceeds,
and the state just after the modular exponentiation processes,
\begin{equation}
|{\rm ME}\rangle
\equiv 
\frac{1}{\sqrt{2^{N_1}}}\sum_{a=0}^{2^{N_1}-1}
|a \rangle_1
|{\sf x}^a \mbox{ mod } M \rangle_2,
\label{psiME}\end{equation}
is a macroscopically entangled state.
Here,  
$| \cdots  \rangle_1$ ($| \cdots  \rangle_2$)
represents a state of the first (second) register, 
${N_1}$ ($2 \log M \leq N_1 < 2 \log M + 1$) 
denotes the number of qubits 
in the first register,
${\sf x}$ is a randomly taken integer,
and $M$ is a large integer to be factored.
This state was shown to play an essential role in 
Shor's factoring algorithm \cite{Ukena}.
Although {\em presence} of entanglement in this state
is obvious, 
the presence of {\em macroscopic} entanglement 
was not revealed until an additive operator 
whose fluctuation is $O(N^2)$ was found in Ref.~\cite{Ukena}.

Another example is entanglement of ground states of 
antiferromagnets, which 
has recently been studied by many authors using various measures 
\cite{Arnesen,Wang,Vidal,O'Connor,Subrahmanyam}.
It is well-known that the {\em exact} ground state
$|{\rm G_{AF}} \rangle$
of the Heisenberg antiferromagnet on a two-dimensional square lattice 
of a finite size
is {\em not} the N\`eel state 
but the {\em symmetric} state that possesses all the symmetries of the 
Hamiltonian \cite{symmetry}.
We here point out that 
$|{\rm G_{AF}} \rangle$ is 
entangled macroscopically, $p=2$.
In fact, the ground state has a long-range order \cite{Horsch},
\begin{equation}
\langle {\rm G_{AF}}|(\hat{M}_\alpha^{\rm st})^2|{\rm G_{AF}} \rangle
\sim0.117N^2+1.02N^{\frac{4}{3}},
\end{equation}
where $\hat{M}_\alpha^{\rm st}\equiv\sum_{l=1}^N(-1)^l\hat{\sigma}_\alpha(l)$ 
is the staggered magnetization $(\alpha = x,y,z)$.
On the other hand, 
$\langle {\rm G_{AF}}|\hat{M}_\alpha^{\rm st}|{\rm G_{AF}} \rangle=0$
by symmetry.
Therefore, 
the order parameter $\hat{M}^{\rm st}_\alpha$ 
of the antiferromagnetic phase transition 
exhibits a huge fluctuation,
\begin{equation}
\langle {\rm G_{AF}} |
(\Delta \hat{M}^{\rm st}_\alpha )^2
|{\rm G_{AF}} \rangle
= O(N^2).
\end{equation}
This shows that $p=2$ for $|{\rm G_{AF}} \rangle$.
Note that such a macroscopically entangled ground state appears 
generally in a {\em finite} system that 
will exhibit a phase transition as $N \to \infty$ 
if the order parameter
does not commute with the Hamiltonian \cite{SM,KT}.
For example, 
the ground state of interacting bosons \cite{pre01,jpsj02},
for which the order parameter is the field operator of the bosons, 
is entangled macroscopically. 
Moreover, the ground state of the transverse Ising model,
whose entanglement has recently been studied using various measures 
\cite{Syljuasen,J.Vidal,Verstraete,Osborne,Vidal}, also has $p=2$
when the transverse magnetic field is below the critical point. 

As demonstrated by these examples, 
the index $p$ captures the presence or absence of
certain anomalous features, which are sometimes hard to 
find intuitively, of 
pure quantum states in finite macroscopic systems.
Furthermore, as we will explain in Sec.~\ref{sec-stability},
$p$ is directly related to fundamental stabilities of many-body states.

\section{macroscopic entanglement of $m$-magnon states}\label{many-partite}
We now study macroscopic entanglement of magnon states (\ref{m-magnon state}) 
with various densities and wavenumbers by evaluating the index p.
\subsection{States to be studied}\label{ss-abc}
Most relevant parameters characterizing the magnon states are 
the number of magnons, $m$, and the wavenumbers of magnons. 
Because of the $Z_2$ symmetry, we assume that $1 \leq m \leq N/2$
without loss of generality.
Furthermore, we assume that $N$ is even in order to avoid uninteresting
complications.

Since we are interested in the asymptotic behavior 
for large $N$, only the order of magnitudes of these parameters
is important.
We therefore consider the following three cases \cite{beyondscope}:
\begin{itemize}
\item[(a)] 
$m=O(1)$.

\item[(b)] 
$m=O(N)$ and all magnons have different wavenumbers 
from each other, continuously occupying a part of 
the first Brillouin zone.
Because of the translational invariance of the system in the $k$-space, 
it is sufficient to calculate the case where 
the magnons continuously occupy 
the first Brillouin zone {\em from the bottom},
i.e., their wavenumbers are 
$0,\pm\frac{2\pi}{N}, \pm\frac{4\pi}{N}, \cdots$, respectively.

\item[(c)]
$m=O(N)$ and all magnons have equal wavenumbers $k$.
Because of the translational 
invariance of the system in the $k$-space, 
we can take $k=0$ without loss of generality. 
\end{itemize}
Furthermore, 
a small number $(=O(1))$ of magnons 
with arbitrary wavenumbers may be added to 
these states.
It is expected and will be confirmed in the following that 
the addition does not alter the value of $p$.

\subsection{Case (a)}\label{sec-me-a}
\begin{figure}[htbp]
\includegraphics[width=0.65\textwidth]{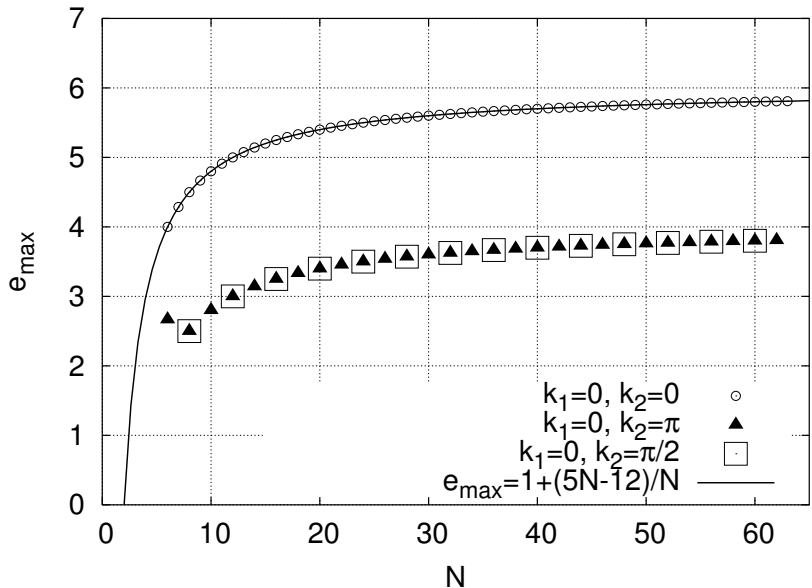}
\caption{The maximum eigenvalue $e_{\rm max}$ 
of the VCM of two-magnon states with wavenumbers $k_1$ and $k_2$
as functions of the number $N$ of spins.
Because of the translational invariance 
of the system in the $k$-space, we 
take $k_1=0$ without loss of generality. 
The solid line represents the analytic expression $e_{{\rm max}}=1+(5N-12)/N$, 
Eq.~(\ref{e_max}), 
which assumes that all wavenumbers are equal.}
\label{fig1.eps}
\end{figure}

In  Fig.~\ref{fig1.eps} we plot numerical results for 
$e_{\rm max}$ of two-magnon states as functions of $N$.
The result for $k_1 = k_2$ can also be obtained from 
the analytic expression, Eq.~(\ref{e_max}).
These results show that 
excitation of a small number ($O(1)$) of magnons 
on the ferromagnetic ground state $|\downarrow^{\otimes N}\rangle$, 
which is a separable state, 
does not change the value of $p$.
It is thus concluded that magnon states for case (a) are 
{\em not} macroscopically entangled.

\subsection{Case (b)}\label{sec-me-b}
\begin{figure}[htbp]
\includegraphics[width=0.65\textwidth]{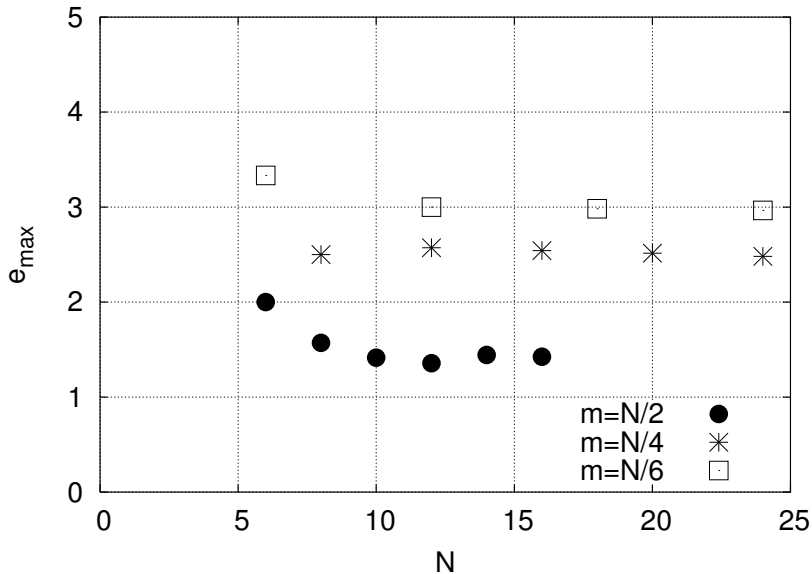}
\caption{The maximum eigenvalue $e_{\rm max}$ of the VCM 
of $m$-magnon states with $m=N/2, N/4$, and $N/6$ 
as functions of the number $N$ of spins.
The wavenumbers of magnons are 
all different taking the values 
$0,\pm2\pi/N,\pm4\pi/N,\ldots$, respectively,
i.e., the first Brillouin zone is continuously occupied from the bottom.}
\label{fig2.eps}
\end{figure}

To investigate $p$ for case (b), 
we evaluate $e_{\rm max}$ for various magnon densities
assuming that all magnons have different wavenumbers from each other,
continuously occupying the first Brillouin zone from the bottom.
The results are plotted in Fig.~\ref{fig2.eps} 
for $m=N/2, N/4$, and $N/6$ as functions of $N$. 
It is seen that $e_{\rm max} \sim$ constant, hence $p=1$.
We also confirmed (not shown in the figure) 
that addition of small number of magnons
with arbitrary wavenumbers does not alter the value of $p$.
We thus conclude that magnon states for case (b) are {\em not} 
macroscopically entangled.

\subsection{Case (c)}\label{sec-me-c}
If the wavenumbers of all magnons are equal, we can 
calculate $e_{\rm max}$ analytically 
as follows. 
Since we can take $k=0$ by symmetry, 
we calculate the VCM of $|\psi_{(k=0)^m;N}\rangle$.
From calculations described in Appendix A, 
we obtain the VCM and the maximum eigenvalue as
\begin{eqnarray}
V_{j,j'}&=&
\left\{
\begin{array}{ll}
1       & (j=j',\ 1\le j \le 2N)\\
1-W_3^2 & (j=j',\ 2N+1\le j \le 3N)\\
W_1     & (j\neq j',\ 1\le j,j'\le N)\\
W_1     & (j\neq j',\ N+1\le j,j'\le 2N)\\
W_2-W_3^2     & (j\neq j',\ 2N+1\le j,j'\le 3N)\\
-iW_3   & (j=j'-N,\ N+1\le j'\le 2N)\\
iW_3    & (j=j'+N,\ 1\le j'\le N)\\
0       & (\mbox{others}),
\end{array}
\right.
\label{Vij}
\end{eqnarray}
\begin{eqnarray}
e_{\rm max}
&=&1+(N-1)W_1+W_3\nonumber\\
&=&1+\frac{2mN-2m^2+N-2m}{N},
\label{e_max}
\end{eqnarray}
where $W_1$, $W_2$, and $W_3$ are defined by 
Eqs.~(\ref{W_1}), (\ref{W_2}), and (\ref{W_3}), respectively.
We therefore find that 
$e_{\rm max} = O(N)$ for $m=O(N)$, hence $p=2$.

\begin{figure}[htbp]
\includegraphics[width=0.65\textwidth]{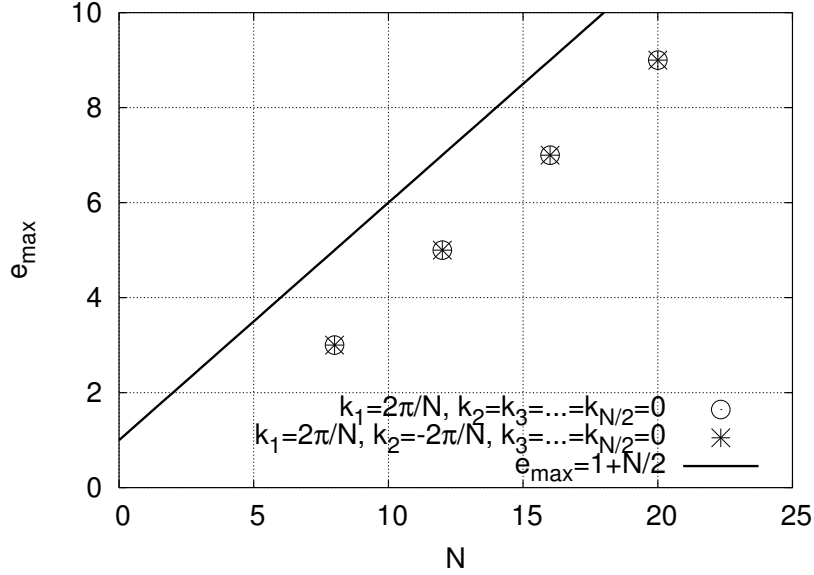}
\caption{The maximum eigenvalue $e_{\rm max}$ of the VCM 
of $N/2$-magnon states 
as functions of the number $N$ of spins. 
Most magnons have equal wavenumbers, i.e, most magnons are 
`condensed.'
The solid line represents the analytic expression 
$e_{\rm max}=1+N/2$, Eq.~(\ref{e_max}), which assumes that 
all wavenumbers are equal.
The circle and crosse represent numerical results for 
the cases where the wavenumbers of one or two magnons are different.
Because of the translational invariance in the 
$k$-space, we take the wavenumber of the condensed magnons as 0 without 
loss of generality.
}
\label{fig3.eps}
\end{figure}

The solid line in Fig.~\ref{fig3.eps} represents  
the analytic expression for $e_{\rm max}$, Eq.~(\ref{e_max}), for $m=N/2$.
We also plot numerical results for 
the cases where the wavenumbers of one or two magnons are different.
It is seen that $e_{\rm max}$ becomes smaller in the latter cases,
as we have seen a similar tendency in Fig.~\ref{fig1.eps}.
However, $e_{\rm max}=O(N)$ and thus $p=2$ 
in all three cases in Fig.~\ref{fig3.eps}.
We therefore conclude that magnon states
for case (c) are macroscopically entangled.

\subsection{Additive operator with the maximum fluctuation}
\label{hermitian}

For a given state $| \psi \rangle$,
we can obtain the additive operator $\hat A_{\rm max}$ 
that has the maximum fluctuation 
$(\delta A_{\rm max})^2 \equiv
\langle \psi | \Delta \hat A^\dagger_{\rm max} \Delta \hat A_{\rm max}
| \psi \rangle =N e_{\rm max}$
for that state by inserting 
the eigenvector of the VCM belonging to the maximum eigenvalue $e_{\rm max}$
into Eq.~(\ref{aaa}).
However, $\hat A_{\rm max}$
is generally non-hermitian
because the eigenvector is generally complex.
A non-hermitian operator $\hat{A}$ can be decomposed into 
the sum of two hermitian operators $\hat A'$ and $\hat A''$ as
$\hat{A} = \hat A' + i \hat A''$.
If $\hat A'$ and $\hat A''$ commute with each other, 
they
can be measured simultaneously with vanishing errors.
Since the values of $\hat{A}$ have one to one correspondence to 
the pairs of the values of $\hat A'$ and $\hat A''$, 
one can measure $\hat{A}$ 
by simultaneously measuring $\hat A'$ and $\hat A''$
if $[\hat A', \hat A'']=0$.
Note that  in macroscopic systems
$[\hat A', \hat A''] \simeq 0$ 
to a good approximation for {\em any} additive operators
$\hat A'$ and $\hat A''$ 
because $[(\hat A'/N), (\hat A''/N)]$ is at most $O(1/N) \simeq 0$.
Therefore, in macroscopic systems
non-hermitian additive operators can be measured to a good accuracy.
Hence, $\hat A_{\rm max}$ can be measured even if it is non-hermitian.
One can also construct the hermitian additive operators 
$\hat A'_{\rm max} \equiv (\hat A_{\rm max}+\hat A^\dagger_{\rm max})/2$
and
$\hat A''_{\rm max} \equiv (\hat A_{\rm max}-\hat A^\dagger_{\rm max})/2i$,
which are the `real' and `imaginary' parts, respectively, 
of $\hat A_{\rm max}$.
Using the triangle inequality 
$
\| \Delta \hat A_{\rm max} | \psi \rangle \|
\leq
\| \Delta \hat A'_{\rm max} | \psi \rangle \|
+\| \Delta \hat A''_{\rm max} | \psi \rangle \|
$,
we can easily show that 
either (or both) of 
$\delta A'_{\rm max}$ or $\delta A''_{\rm max}$
is of the same order as $\delta A_{\rm max}$.

For $|\psi_{(k=0)^m;N}\rangle$ with $m=O(N)$, for example, 
the eigenvector belonging to the maximum eigenvalue (\ref{e_max}) is
\begin{equation}
\frac{1}{2}(\displaystyle\overbrace{1,\ldots,1}^N,
\displaystyle\overbrace{i,\ldots,i}^N,
\displaystyle\overbrace{0,\ldots,0}^N)^t,
\label{evec}
\end{equation}
which gives the maximally fluctuating additive operator as 
\begin{equation}
\hat A_{\rm max}
=
\frac{1}{2}\sum_{l=1}^N\left(1\cdot\hat{\sigma}_x(l)+
i\cdot\hat{\sigma}_y(l)+
0\cdot\hat{\sigma}_z(l)\right)
=\sum_{l=1}^N\hat{\sigma}_+(l),
\end{equation}
for which $(\delta A_{\rm max})^2 = O(N^2)$.
Although this operator is not hermitian, it can be measured to a good accuracy
if $N \gg 1$.
Or, let us define hermitian operators 
$\hat A'_{\rm max} \equiv (\hat A_{\rm max}+\hat A^\dagger_{\rm max})/2
=\frac{1}{2}\sum_{l=1}^N\hat{\sigma}_x(l)$
and
$\hat A''_{\rm max} \equiv (\hat A_{\rm max}-\hat A^\dagger_{\rm max})/2i
=\frac{1}{2}\sum_{l=1}^N\hat{\sigma}_y(l)$.
Since $|\psi_{(k=0)^m;N}\rangle$ is symmetric under rotations 
about the $z$ axis, 
we can show that 
$(\delta A'_{\rm max})^2 = (\delta A''_{\rm max})^2 = O(N^2)$ in this case.
It is worth mentioning that 
$\hat A^\dagger_{\rm max}$ corresponds to 
the eigenvector belonging to the second largest eigenvalue 
$e_4$, which is given by Eq.\ (\ref{lambda4}) and is of $O(N)$.

\section{bipartite entanglement of $m$-magnon states}\label{two-partite}
For a comparison purpose, 
we now calculate the degree of bipartite entanglement
of magnon states that have been studied in the previous section.
For a measure of bipartite entanglement, we use the von Neumann entropy 
of the reduced density operator of a subsystem. 
That is, we halve the $N$-spin system 
and evaluate the reduced density operator 
$\hat \rho_{N/2}(N)$ of one of the halves.
The von Neumann entropy is defined by 
\begin{equation}
S_{N/2}(N) \equiv -{\rm Tr}\left[ 
\hat \rho_{N/2}(N) \log_2 \hat \rho_{N/2}(N)\right],
\label{SN/2}\end{equation}
which ranges from $0$ to $N/2$.
Although $S_{N/2}(N)$ for 
the case where all 
wavenumbers are equal was discussed by Stockton {\em et al.}\ \cite{Stockton}, 
we here evaluate $S_{N/2}(N)$ systematically for 
all the three cases listed in Sec.~\ref{ss-abc}.

\subsection{Case (a)}
To evaluate $S_{N/2}(N)$, we halve the $N$-spin system into 
two subsystems A and B.
Accordingly, we decompose 
$|\psi_{k_1,k_2,\ldots,k_m;N}\rangle$ into the sum of products
of $|\psi_{k_1,k_2,\ldots;N/2}\rangle$'s of A and B.

When all wavenumbers are different from each other, an $m$-magnon state
can be decomposed as
\begin{eqnarray}
&& |\psi_{k_1,k_2,\ldots,k_m;N}\rangle
=
G_{k_1,k_2,\ldots,k_m;N}
\prod_{i=1}^m\hat{M}_{k_i}^{\dagger}
|\downarrow^{\otimes N}\rangle
\nonumber\\
&& =
\frac{G_{k_1,k_2,\ldots,k_m;N}}{\sqrt{2^m}}
\left(G_{k_1,\cdots,k_m;N/2}^{-1}
|\downarrow^{\otimes N/2}\rangle
|\psi_{k_1,\ldots,k_m;N/2}\rangle\right.\nonumber\\
&&{}+\sum_{i=1}^me^{ik_iN/2}
G_{k_i;N/2}^{-1}G_{k_1,\cdots,\tilde{k_i},\cdots,k_m;N/2}^{-1}
|\psi_{k_i;N/2}\rangle
|\psi_{k_1,\ldots,\tilde{k}_i,\ldots,k_m;N/2}\rangle\nonumber\\
&&{}+\sum_{i=1}^{m-1}\sum_{j=i+1}^me^{ik_iN/2+ik_jN/2}
G_{k_i,k_j;N/2}^{-1}
G_{k_1,\cdots,\tilde{k_i},\cdots,\tilde{k_j},\cdots,k_m;N/2}^{-1}
|\psi_{k_i,k_j;N/2}\rangle
|\psi_{k_1,\ldots,\tilde{k}_i,\ldots,\tilde{k}_j,\ldots,k_m;N/2}\rangle
\nonumber\\
&&{} \quad \vdots\nonumber\\
&&{}+\left.e^{ik_1N/2+\ldots+ik_mN/2}
G_{k_1,\cdots,k_m;N/2}^{-1}
|\psi_{k_1,\ldots,k_m;N/2}\rangle
|\downarrow^{\otimes N/2}\rangle
\right),
\label{decomposition}
\end{eqnarray}
where 
$\tilde{\quad}$ denotes absence, 
and the prefactor $1/\sqrt{2^m}$ comes from 
the prefactor $1/\sqrt{N}$ in Eq.~(\ref{Mdagger}).
When $m=2$, for example, 
\begin{eqnarray}
|\psi_{k_1,k_2;N}\rangle
&=&
\frac{G_{k_1,k_2;N}}{2}\left(G_{k_1,k_2;N/2}^{-1}
|\downarrow^{\otimes N/2}\rangle
|\psi_{k_1,k_2;N/2}\rangle\right.\nonumber\\
{}&&+e^{ik_1N/2}G_{k_1;N/2}^{-1}G_{k_2;N/2}^{-1}
|\psi_{k_1;N/2}\rangle
|\psi_{k_2;N/2}\rangle
+e^{ik_2N/2}G_{k_1;N/2}^{-1}G_{k_2;N/2}^{-1}
|\psi_{k_2;N/2}\rangle
|\psi_{k_1;N/2}\rangle\nonumber\\
{}&&+\left.e^{ik_1N/2+ik_2N/2}G_{k_1,k_2;N/2}^{-1}
|\psi_{k_1,k_2;N/2}\rangle
|\downarrow^{\otimes N/2}\rangle
\right),
\label{4terms}
\end{eqnarray}
which means that the state is a superposition of the following 
four ($=2^2$) states: 
(i) both magnons are in subsystem $B$, 
(ii) the magnon with $k_1$ is in $A$ whereas the magnon with $k_2$ is in $B$, 
(iii) the magnon with $k_2$ is in $A$ whereas the magnon with $k_1$ is in $B$, 
and 
(iv) both magnons are in A. 
As $N$ is increased in decomposition 
(\ref{decomposition}) (while $m$ is fixed), 
all $G$'s $\to 1$ and
$2^m$ vectors on the right-hand side
tend to become orthonormalized.
This means that decomposition (\ref{decomposition}) becomes the 
Schmidt decomposition, in which the Schmidt rank is $2^m$
and all the Schmidt coefficients are equal 
(except for the phase factors).
We thus obtain
\begin{equation}
\lim_{N \to \infty \atop \mbox{($m$: fixed)}}
S_{N/2}(N)
=
-\sum_{i=1}^{2^m}\left(\frac{1}{\sqrt{2^m}}\right)^2
\log_2\left(\frac{1}{\sqrt{2^m}}\right)^2
=m,
\label{S-case(a)-1}\end{equation}
i.e., 
$S_{N/2}(N) = O(1)$ for fixed $m$.
Note that $m$ is the maximum value of
$S_{N/2}(N)$ among states whose Schmidt rank is $2^m$.

When some of the wavenumbers are equal, 
the Schmidt rank 
becomes smaller 
because magnons having equal wavenumbers are 
indistinguishable.
For example, if $k_1=k_2\equiv k$, the two-magnon 
state
\begin{eqnarray}
|\psi_{k,k;N}\rangle
=\frac{G_{k,k;N}}{\sqrt{2}}
(\hat{M}_k^\dagger)^2
|\downarrow^{\otimes N}\rangle
\end{eqnarray}
is decomposed as
\begin{eqnarray}
|\psi_{k,k;N}\rangle&=&\frac{G_{k,k;N}}{2}\left(G_{k,k;N/2}^{-1}
|\downarrow^{\otimes N/2}\rangle
|\psi_{k,k;N/2}\rangle\right.\nonumber\\
{}&&+\sqrt{2}e^{ikN/2}G_{k;N/2}^{-1}G_{k;N/2}^{-1}
|\psi_{k;N/2}\rangle
|\psi_{k;N/2}\rangle\nonumber\\
{}&&+\left.e^{ikN}G_{k,k;N/2}^{-1}
|\psi_{k,k;N/2}\rangle
|\downarrow^{\otimes N/2}\rangle
\right).
\end{eqnarray}
In contrast to Eq.~(\ref{4terms}), 
this is the superposition of the following {\em three} 
($< 2^2$) states; 
(i) both magnons are in $B$, 
(ii) one magnon is in $A$ and the other is in $B$, 
and (iii) both magnons are in $A$.
The Schmidt rank is thus decreased.
Furthermore, 
the Schmidt coefficients do not take the same value.
As a result of these, 
$S_{N/2}(N)$ becomes smaller than Eq.~(\ref{S-case(a)-1});
\begin{equation}
\lim_{N \to \infty \atop \mbox{($m$: fixed)}}
S_{N/2}(N)
<
m,
\label{S-case(a)-2}\end{equation}
from which we again have 
$S_{N/2}(N) = O(1)$. 

Therefore, 
we conclude that 
the bipartite entanglement of magnon states 
in case (a) is small in the sense that
\begin{eqnarray}
S_{N/2}(N)=O(1).
\label{S-a=1}\end{eqnarray}

\begin{figure}[htbp]
\includegraphics[width=0.65\textwidth]{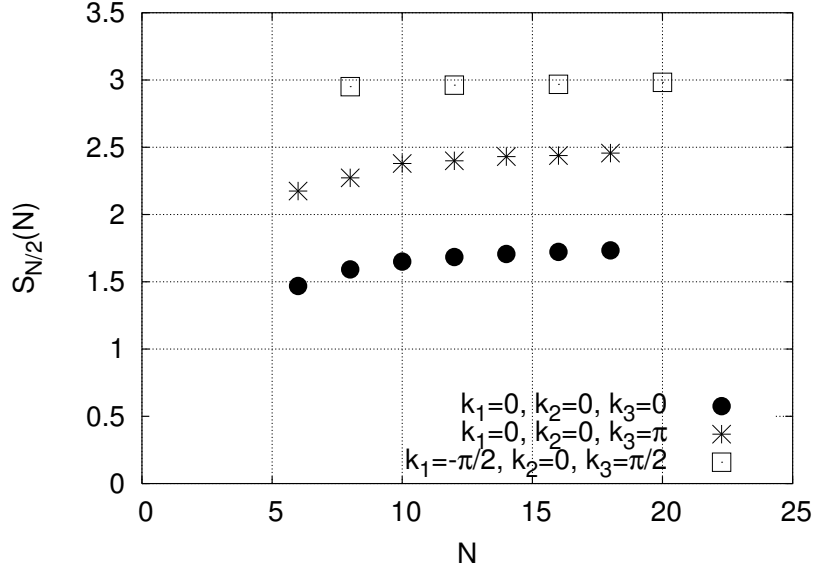}
\caption{The von Neumann entropy of a subsystem, $S_{N/2}(N)$, 
of three-magnon states with wavenumbers $k_1$, $k_2$, and $k_3$ 
as functions of the number $N$ of spins. 
Because of the translational invariance of 
the system in the $k$-space, we take $k_2=0$ 
without loss of generality.}
\label{fig4.eps}
\end{figure}

As a demonstration, 
we plot numerical results for $S_{N/2}(N)$
as functions of $N$ in Fig.~\ref{fig4.eps}, 
for three-magnon states in the following three cases;
(i) three magnons have different wavenumbers $k_1, k_2, k_3$, 
(ii) two magnons have equal wavenumbers $k_1=k_2$ whereas one magnon 
has another wavenumber $k_3$, 
and 
(iii) three magnons have equal wavenumbers $k_1 = k_2 =k_3$.
Formulas (\ref{S-case(a)-1}) and (\ref{S-case(a)-2})
are confirmed.
Furthermore, it is seen that 
the departure of magnons from 
ideal bosons becomes significant for small $N$, 
and that $S_{N/2}(N)$ approaches 
the limiting values for $N\to\infty$
from below.
This may be understood from the discussions of the following subsections.

Note that the result (\ref{S-a=1})
agrees in some sense with the result of Sec.~\ref{sec-me-a}, in which 
we have seen that the states are not macroscopically 
entangled.
However, we will see in the following that 
such a simple agreement is not obtained in cases (b) and (c).

\subsection{Case (b)}
\label{ss-case(b)}

In case (b), the previous argument on 
$\lim_{N \to \infty} S_{N/2}(N)$ does not hold 
because the departure of magnons from 
ideal bosons is significant when $m = O(N)$.
In fact, the vectors in Eq.~(\ref{decomposition}) do not become
orthonormalized as $N \to \infty$.
Hence, Eq.~(\ref{decomposition}) does not become the Schmidt decomposition, 
and it can be 
further arranged 
until 
it becomes the Schmidt decomposition.
Therefore,  we expect that the Schmidt rank 
is less than $2^m$, and $\lim_{N \to \infty} S_{N/2}(N)<m$.

\begin{figure}[htbp]
\includegraphics[width=0.65\textwidth]{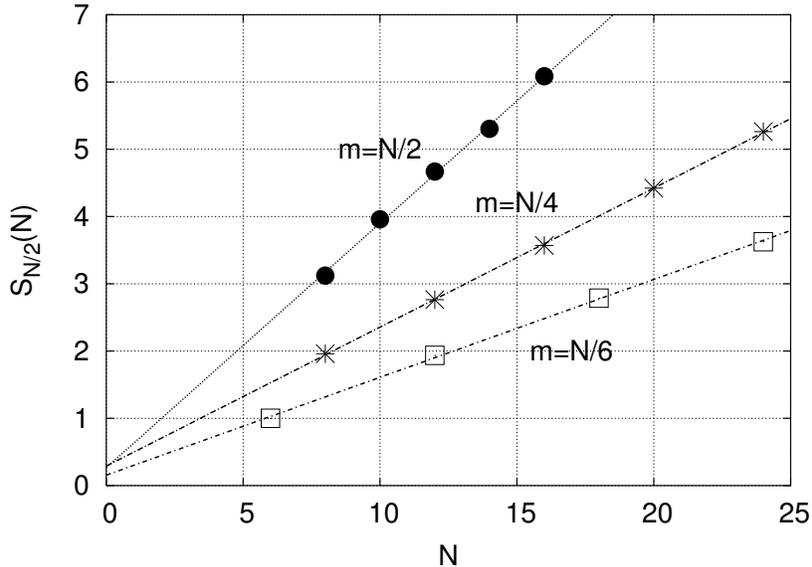}
\caption{The von Neumann entropy of a subsystem, $S_{N/2}(N)$, 
of $m$-magnon states with $m=N/2, N/4$, and $N/6$ 
as functions of the number $N$ of spins.
The wavenumbers of magnons are all different taking the values 
$0,\pm2\pi/N,\pm4\pi/N,\ldots$, respectively, 
i.e., the first Brillouin zone is continuously occupied from 
the bottom. 
The lines represent the regression lines calculated with the least squares.
}
\label{fig5.eps}
\end{figure}

To see more details, 
we have calculated $S_{N/2}(N)$ numerically.
The results are plotted as functions of $N$ in 
Fig.~\ref{fig5.eps}
for case (b) with $m=N/2, N/4$, and $N/6$.
It is found that the results are well approximated by
the straight lines, 
\begin{equation}
S_{N/2}(N) = aN+b, 
\label{rl}\end{equation}
which are also displayed in Fig.~\ref{fig5.eps}.
The parameters $a$ and $b$ are determined by the least squares, 
whose values are tabulated in Table~\ref{table}.
Since $0<a<m/N$, we find that 
$S_{N/2}(N)$ is less than, but of the same order of magnitude as,
the maximum value $N/2$;
\begin{eqnarray}
S_{N/2}(N)=O(N).
\label{S=O(N)}\end{eqnarray}
We thus conclude that 
the bipartite entanglement of magnon states in case (b)
is extremely large. 
This should be contrasted with the result of Sec.~\ref{sec-me-b}, 
according to which these states are {\em not} macroscopically entangled.

\begin{table}
\caption{The parameters $a$ and $b$, which are 
calculated with the least squares, 
of the regression line Eq.~(\ref{rl}) for $m$-magnon states
of Fig.~\ref{fig5.eps}.
}
\begin{ruledtabular}
\begin{tabular}{ccccc}
$m$& a & asymptotic standard error & b & 
asymptotic standard error\\
\hline
N/2 & 0.36 & $\pm$ 0.009 & 0.27 & $\pm$ 0.113\\
N/4 & 0.21 & $\pm$ 0.002 & 0.29 & $\pm$ 0.027\\
N/6 & 0.15 & $\pm$ 0.003 & 0.15 & $\pm$ 0.044\\
\end{tabular}
\end{ruledtabular}
\label{table}
\end{table}

\subsection{Case (c)}
\label{ss-case(c)}
We finally consider $S_{N/2}(N)$ in case (c).
When all wavenumbers are equal to zero,
the $m$-magnon state $|\psi_{(k=0)^m;N}\rangle$
becomes identical to the ``Dicke state'' 
that was discussed by Stockton {\em et al.} \cite{Stockton}. 
According to their result, 
\begin{eqnarray}
S_{N/2}(N)=O(\log N)
\label{S=O(logN)}
\end{eqnarray}
when $m=O(N)$. 
Because of the translational invariance,
this result also holds for $|\psi_{(k)^m;N}\rangle$ with other values of $k$.
Since $1 \lesssim O(\log N) \ll O(N)$, 
we find  that 
$S_{N/2}(N)$ is slightly larger than that of case (a), 
but much smaller than that of case (b).
We therefore conclude that 
the bipartite entanglement of magnon states for case (c)
is small.
This should be contrasted with the result of Sec.~\ref{sec-me-c}, 
according to which these states are macroscopically entangled.

\section{Stabilities and entanglement}\label{sec-stability}
It may be expected that a quantum state 
with larger entanglement would be more unstable.
This naive expectation is, however, quite ambiguous for 
many-body systems.
First of all, 
the degree of entanglement depends drastically 
on the measure or index used to quantify the entanglement,
as we have shown above.
Furthermore, 
`stability' can be defined in many different ways for many-body states.

SM considered the following two kinds of stabilities \cite{SM}.
One is the stability against weak perturbations from 
noises or environments:
A pure state is said to be {\em fragile} 
if its decoherence rate behaves as $\sim K N^{1+\delta}$
when perturbations from the noises or environments are weak, 
where $\delta$ is a positive constant.
Such a state is extremely unstable in the sense that 
its decoherence rate {\em per spin} increases as $\sim K N^{\delta}$
with increasing $N$, until it becomes extremely large 
for huge $N$ however small is the coupling constant 
between the system and the noise or environment. 
SM showed that pure states with $p=1$ never become fragile 
in any noises or environments, whereas 
pure states with $p=2$ 
can become fragile, depending on the spectral intensities 
of the noise or environment variables. 
The other stability considered by SM is 
the stability against local measurements:
A state is said to be 
{\em stable against local measurements} if an ideal (projective) 
measurement of any observable at a point $l$ does not 
alter the result of measurement of any observable at a
distant point $l'$ for sufficiently large $|l-l'|$. 
SM showed that 
this stability is equivalent to the `cluster property', which is 
closely related to $p$, 
if the cluster property for finite systems is properly defined. 
For example, a state is {\em unstable} against local measurements
if $p=2$, whereas 
a homogeneous \cite{homogeneous}
state with $p=1$ 
is stable.

We have shown 
that $p=1$ for magnon states
of case (b). 
Therefore, these states never become fragile in any noises or environments, 
and they are stable against local measurements. 
Since we have also shown that 
the bipartite entanglement of these states is 
extremely large,  
we find that the bipartite entanglement is basically 
independent of these fundamental stabilities.
The same conclusion was obtained for chaotic quantum systems by 
two of the authors \cite{Sugita}:
They showed that $p=1$ for almost all energy eigenstates of 
macroscopic chaotic systems  
whereas their bipartite entanglement is nearly maximum. 

We have also shown that $p=2$ for magnon states of case (c).
Therefore, these states 
can become fragile, depending on the spectral intensities 
of the noise or environment variables.
Furthermore, these states are unstable against local measurements.
Since we have seen that 
the bipartite entanglement of these states is small, 
we find again that the bipartite entanglement is basically 
independent of these fundamental stabilities.
To understand the physics of this conclusion, 
the following simple example may be helpful.
The W-state, 
\begin{eqnarray}
|{\rm W} \rangle
=\frac{1}{\sqrt{N}}\sum_{l=1}^N|\downarrow^{\otimes (l-1)}
\uparrow\downarrow^{\otimes(N-l)}\rangle,
\label{W}\end{eqnarray}
has the {\em same} value of $S_{N/2}(N)$
as the $N$-spin GHZ state, i.e., $S_{N/2}(N)=1$.
On the other hand, $p=1$ for $|{\rm W} \rangle$ whereas
$p=2$ for $|{\rm GHZ} \rangle$.
As a result, 
the decoherence rate of $|{\rm W} \rangle$
{\em never} exceeds $O(N)$ by 
{\em any} weak classical noises, 
whereas the decoherence rate of $|{\rm GHZ} \rangle$
becomes as large as $O(N^2)$ in a long-wavelength noise.
Furthermore, $|{\rm W} \rangle$ is {\em stable} against 
local measurements, whereas $|{\rm GHZ} \rangle$
is unstable.
These results are physically reasonable because 
the W-state is nothing but a one-magnon state (with $k=0$), 
which can be generated easily by experiment:
Such a state does not seem very unstable.
(See also Sec.~\ref{sec-origin}.)

It should be mentioned that 
another stability was studied by Stockton {\em et al.}~\cite{Stockton}
for a special state of case (c), i.e., for a 
Dicke state which in our notation 
is written as $|\psi_{(k=0)^{N/2};N}\rangle$.
They showed that a bipartite entanglement measure
of $\hat{\rho}_{N-N'}\equiv Tr_{N'}
(|\psi_{(k=0)^{N/2};N}\rangle \langle\psi_{(k=0)^{N/2};N}|)$ 
decreases very slowly as $N'$ increases. Here,
$Tr_{N'}$ means `trace out $N'$ spins'. 
They thus concluded that the state is {\em robust}.
Although their conclusion might look 
contradictory to our conclusion, 
there is no contradiction.
The stability (robustness) as discussed by 
Stockton {\em et al.} is totally different from the 
fundamental stabilities that are discussed in the present paper.
The Dicke state is `robust' in the sense of Stockton {\em et al.}, 
whereas in the senses of SM 
the state is `fragile' in noises or environments
and `unstable' against local measurements.
This demonstrates that 
stability can be defined in many 
different ways for many-body states.

\section{Discussions}
\subsection{Relation to Bose-Einstein condensates}\label{sec-BEC}
The `cluster property,' which is closely related to the index $p$, 
of condensed states of interacting many bosons was previously studied 
in Ref.~\cite{jpsj02}.
It was shown there that 
$p=2$ for the ground state $|N, {\rm G} \rangle$, 
which has a fixed number $N$ of bosons, 
if  $N$ is large enough to give a finite 
density for a large volume.
Since magnons are approximate bosons, 
magnon states of case (c) may be analogous to this state.
Although deviations from ideal bosons become significant in case (c), 
the deviations may be partly regarded as effective interactions among 
magnons.
This analogy intuitively explains our result that $p=2$ for 
magnon states of case (c).

It was also shown in Ref.~\cite{jpsj02} that $p=1$ for 
a generalized coherent state $|\alpha, {\rm G} \rangle$, 
which was called there
a coherent state of interacting bosons.
This result may also be understood intuitively on the same analogy.
That is, $|\alpha, {\rm G} \rangle$
may be analogous 
to the state of Eq.~(\ref{spon}),
which has $p=1$ because it is a separable state as seen 
from Eq.~(\ref{separable}).
Therefore, by analogy, $p$ should also be unity for 
$|\alpha, {\rm G} \rangle$, 
in consistency with the result of Ref.~\cite{jpsj02},
although $|\alpha, {\rm G} \rangle$ is {\em not} separable.

Analogy like these may be useful for further understanding of 
systems of interacting many bosons and of 
many magnons.
\subsection{What generates huge entanglement?}\label{sec-origin}
We have shown that states with huge entanglement, 
as measured by either $p$ or $S_{N/2}(N)$, 
can be easily constructed by 
simply exciting many magnons on a separable state.
We now discuss the physical origin of this fact.

The most important point is that a magnon propagates spatially 
all over the magnet \cite{propagate}.
By the propagation, quantum coherence is established between 
spatially separated points \cite{propagate2}.
Therefore, by exciting a macroscopic number of magnons, 
one can easily construct states with huge entanglement. 

Note that this 
should be common 
to most quantum systems, 
because Hamiltonians of most physical systems should have 
a term which causes spatial propagation.
For example, such a term includes the nearest-neighbor interaction of
spin systems, the kinetic-energy term 
of the Schr\"odinger equation of particles,
the term composed of spatial derivative of a field operator
in field theory.
Therefore, excitation of a macroscopic number of elementary excitations 
generates huge entanglement. 
Neither randomness nor elaborate tuning is necessary.

This observation will be useful for {\em theoretically} constructing 
states with huge entanglement.
{\em Experimentally}, on the other hand, 
the stability should also be taken into account because
unstable states would be hard to generate experimentally.
We thus consider that states with $p=2$ 
should be much harder to generate experimentally than 
states with large $S_{N/2}(N)$.
In other words, a state with large $S_{N/2}(N)$ would be able to 
be generated rather easily, e.g., by exciting many quasi-particles
in a solid.
In this respect, a naive expectation that states with large entanglement
would be hard to generate experimentally is false: 
It depends on the measure or index that is used to quantify the 
entanglement.
\begin{acknowledgments}
This work is partially supported by Grant-in-Aid for Scientific 
Research.
A. Sugita is supported by Japan Society for the Promotion 
of Science for Young Scientists.
\end{acknowledgments}
\appendix
\section{Calculation of the VCM and its eigenvalues}

The state vector of the $m$-magnon state 
with $k_1 = \cdots = k_m = 0$ can be written as
\begin{eqnarray}
|\psi_{(k=0)^m;N}\rangle=\frac{1}{\sqrt{{}_NC_m}}
\sum_{l_1}\sum_{l_2 \, (>l_1)}\sum_{l_3\, (>l_2)}
\cdots\sum_{l_m\, (>l_{m-1})}
\hat{\sigma}_+(l_1)\hat{\sigma}_+(l_2)\cdots\hat{\sigma}_+(l_m)
|\downarrow^{\otimes N}\rangle,\nonumber
\label{m-magnon}
\end{eqnarray}
where $\displaystyle {}_NC_m \equiv {N \choose m}$.
Since the VCM is hermitian, we have only to calculate the 
following correlations;
$\langle\Delta\hat{\sigma}_x(l)\Delta\hat{\sigma}_x(l')\rangle$,
$\langle\Delta\hat{\sigma}_x(l)\Delta\hat{\sigma}_y(l')\rangle$,
$\langle\Delta\hat{\sigma}_x(l)\Delta\hat{\sigma}_z(l')\rangle$,
$\langle\Delta\hat{\sigma}_y(l)\Delta\hat{\sigma}_y(l')\rangle$,
$\langle\Delta\hat{\sigma}_y(l)\Delta\hat{\sigma}_z(l')\rangle$,
$\langle\Delta\hat{\sigma}_z(l)\Delta\hat{\sigma}_z(l')\rangle$,
where $\langle \cdot \rangle$ stands for 
$\langle\psi_{(k=0)^m;N}|\cdot|\psi_{(k=0)^m;N}\rangle$.
Since $|\psi_{(k=0)^m;N}\rangle$ is an eigenvector of 
$\exp(-i\theta\sum_{i=1}^N\hat{\sigma}_z(l))$, 
the state vector is invariant 
under a rotation about $z$-axis.
Therefore 
$\langle\Delta\hat{\sigma}_y(l)\Delta\hat{\sigma}_y(l')\rangle
=\langle\Delta\hat{\sigma}_x(l)\Delta\hat{\sigma}_x(l')\rangle$ and 
$\langle\Delta\hat{\sigma}_y(l)\Delta\hat{\sigma}_z(l')\rangle
=\langle\Delta\hat{\sigma}_x(l)\Delta\hat{\sigma}_z(l')\rangle$.
Thus we calculate only
\begin{eqnarray}
\langle\Delta\hat{\sigma}_x(l)\Delta\hat{\sigma}_x(l')\rangle,
\langle\Delta\hat{\sigma}_x(l)\Delta\hat{\sigma}_y(l')\rangle,
\langle\Delta\hat{\sigma}_x(l)\Delta\hat{\sigma}_z(l')\rangle,
\langle\Delta\hat{\sigma}_z(l)\Delta\hat{\sigma}_z(l')\rangle.\nonumber
\end{eqnarray}
We note that 
$\langle\hat{\sigma}_x(l)\rangle=\langle\hat{\sigma}_y(l)\rangle=0$
by symmetry, and
\begin{eqnarray}
\langle\hat{\sigma}_z(l)\rangle
=\frac{1}{{}_NC_m}({}_{N-1}C_{m-1}-{}_{N-1}C_m)
=-\frac{N-2m}{N}
\equiv-W_3.\label{W_3}
\end{eqnarray}
When $l=l'$, we easily obtain
$\langle\hat{\sigma}_x(l)\hat{\sigma}_x(l)\rangle=
\langle\hat{\sigma}_z(l)\hat{\sigma}_z(l)\rangle=1$, 
$\langle\hat{\sigma}_x(l)\hat{\sigma}_z(l)\rangle=0$,
and
\begin{eqnarray}
\langle\hat{\sigma}_x(l)\hat{\sigma}_y(l)\rangle&=&
i\langle\hat{\sigma}_z(l)\rangle=-iW_3\nonumber.
\end{eqnarray}
Therefore,
\begin{eqnarray}
\langle\Delta\hat{\sigma}_x(l)\Delta\hat{\sigma}_x(l)\rangle
&=&1,\nonumber\\
\langle\Delta\hat{\sigma}_x(l)\Delta\hat{\sigma}_y(l)\rangle
&=&-iW_3,\nonumber\\
\langle\Delta\hat{\sigma}_x(l)\Delta\hat{\sigma}_z(l)\rangle
&=&0,\nonumber\\
\langle\Delta\hat{\sigma}_z(l)\Delta\hat{\sigma}_z(l)\rangle
&=&1-W_3^2.\nonumber
\end{eqnarray}
When $l\neq l'$, we note that 
\begin{eqnarray}
\sqrt{{}_NC_m}\langle\psi_{(k=0)^m;N}|\hat{\sigma}_x(l)\hat{\sigma}_x(l')
|\ldots\displaystyle\overbrace{\downarrow}^l
\ldots\displaystyle\overbrace{\uparrow}^{l'}\ldots\rangle&=&1,\nonumber\\
\sqrt{{}_NC_m}\langle\psi_{(k=0)^m;N}|\hat{\sigma}_x(l)\hat{\sigma}_x(l')|
\ldots\displaystyle\overbrace{\uparrow}^l
\ldots\displaystyle\overbrace{\downarrow}^{l'}
\ldots\rangle&=&1,\nonumber\\
\sqrt{{}_NC_m}\langle\psi_{(k=0)^m;N}|\hat{\sigma}_x(l)\hat{\sigma}_x(l')|
\ldots\displaystyle\overbrace{\downarrow}^l
\ldots\displaystyle\overbrace{\downarrow}^{l'}\ldots\rangle&=&0,\nonumber\\
\sqrt{{}_NC_m}\langle\psi_{(k=0)^m;N}|\hat{\sigma}_x(l)\hat{\sigma}_x(l')|
\ldots\displaystyle\overbrace{\uparrow}^l
\ldots\displaystyle\overbrace{\uparrow}^{l'}\ldots\rangle&=&0,\nonumber
\end{eqnarray}
where 
$|\ldots\displaystyle\overbrace{\downarrow}^l
\ldots\displaystyle\overbrace{\uparrow}^{l'}\ldots\rangle$
is a state vector in which $m$ spins including 
$l'$-th spin are up, whereas $N-m$ spins 
including $l$-th spin are down.
We thus obtain
\begin{eqnarray}
\langle\Delta\hat{\sigma}_x(l)\Delta\hat{\sigma}_x(l')\rangle
=\frac{2{}_{N-2}C_{m-1}}{{}_NC_m}
=\frac{2m(N-m)}{N(N-1)}
\equiv W_1.\label{W_1}
\end{eqnarray}
Furthermore, since 
\begin{eqnarray}
\sqrt{{}_NC_m}\langle\psi_{(k=0)^m;N}|\hat{\sigma}_x(l)\hat{\sigma}_y(l')
|\ldots\displaystyle\overbrace{\downarrow}^l
\ldots\displaystyle\overbrace{\uparrow}^{l'}\ldots\rangle&=&i,\nonumber\\
\sqrt{{}_NC_m}\langle\psi_{(k=0)^m;N}|\hat{\sigma}_x(l)\hat{\sigma}_y(l')|
\ldots\displaystyle\overbrace{\uparrow}^l
\ldots\displaystyle\overbrace{\downarrow}^{l'}\ldots\rangle&=&-i,\nonumber\\
\sqrt{{}_NC_m}\langle\psi_{(k=0)^m;N}|\hat{\sigma}_x(l)\hat{\sigma}_y(l')|
\ldots\displaystyle\overbrace{\downarrow}^l
\ldots\displaystyle\overbrace{\downarrow}^{l'}\ldots\rangle&=&0,\nonumber\\
\sqrt{{}_NC_m}\langle\psi_{(k=0)^m;N}|\hat{\sigma}_x(l)\hat{\sigma}_y(l')|
\ldots\displaystyle\overbrace{\uparrow}^l
\ldots\displaystyle\overbrace{\uparrow}^{l'}\ldots\rangle&=&0,\nonumber
\end{eqnarray}
we obtain
\begin{eqnarray}
\langle\Delta\hat{\sigma}_x(l)\Delta\hat{\sigma}_y(l')\rangle
=\frac{1}{{}_NC_m}\left(i{}_{N-2}C_{m-1}-i{}_{N-2}C_{m-1}\right)\nonumber
=0.\nonumber
\end{eqnarray}
It is obvious that
$
\langle\hat{\sigma}_x(l)\hat{\sigma}_z(l')\rangle=0
$
because $|\psi_{(k=0)^m;N}\rangle$ is a linear combination of vectors whose 
$m$ spins are up and $N-m$ spins are down.
Therefore 
\begin{eqnarray}
\langle\Delta\hat{\sigma}_x(l)\Delta\hat{\sigma}_z(l')\rangle=0,\nonumber
\end{eqnarray}
Finally, since
\begin{eqnarray}
\sqrt{{}_NC_m}\langle\psi_{(k=0)^m;N}|\hat{\sigma}_z(l)\hat{\sigma}_z(l')|
\ldots\displaystyle\overbrace{\downarrow}^l
\ldots\displaystyle\overbrace{\uparrow}^{l'}\ldots\rangle&=&-1,\nonumber\\
\sqrt{{}_NC_m}\langle\psi_{(k=0)^m;N}|\hat{\sigma}_z(l)\hat{\sigma}_z(l')|
\ldots\displaystyle\overbrace{\uparrow}^l
\ldots\displaystyle\overbrace{\downarrow}^{l'}\ldots\rangle&=&-1,\nonumber\\
\sqrt{{}_NC_m}\langle\psi_{(k=0)^m;N}|\hat{\sigma}_z(l)\hat{\sigma}_z(l')|
\ldots\displaystyle\overbrace{\downarrow}^l
\ldots\displaystyle\overbrace{\downarrow}^{l'}\ldots\rangle&=&1,\nonumber\\
\sqrt{{}_NC_m}\langle\psi_{(k=0)^m;N}|\hat{\sigma}_z(l)\hat{\sigma}_z(l')|
\ldots\displaystyle\overbrace{\uparrow}^l
\ldots\displaystyle\overbrace{\uparrow}^{l'}\ldots\rangle&=&1,\nonumber
\end{eqnarray}
we obtain
\begin{eqnarray}
\langle\hat{\sigma}_z(l)\hat{\sigma}_z(l')\rangle
&=&\frac{1}{{}_NC_m}\left(-2{}_{N-2}C_{m-1}
+{}_{N-2}C_m+{}_{N-2}C_{m-2}\right)\nonumber\\
&=&\frac{N^2-4mN-N+4m^2}{N(N-1)}\nonumber\\
&\equiv&W_2.\label{W_2}
\end{eqnarray}
Therefore,
\begin{eqnarray}
\langle\Delta\hat{\sigma}_z(l)\Delta\hat{\sigma}_z(l')\rangle
=W_2-W_3^2.\nonumber
\end{eqnarray}
Combining these results, we obtain the VCM as Eq.~(\ref{Vij}), or
\begin{eqnarray}
V=
\left(
\begin{array}{ccc|ccc|ccc}
1   &      &W_1 &-iW_3&      &0    &0        &      &         \\
    &\ddots&    &     &\ddots&     &         &\ddots&         \\ 
W_1 &      &1   &0    &      &-iW_3&         &      &0        \\ 
\hline
iW_3&      &0   &1    &      &W_1  &0        &      &         \\
    &\ddots&    &     &\ddots&     &         &\ddots&         \\
0   &      &iW_3&W_1  &      &1    &         &      &0        \\
\hline
0   &      &    &0    &      &     &1-W_3^2  &      &W_2-W_3^2\\
    &\ddots&    &     &\ddots&     &         &\ddots&         \\
    &      &0   &     &      &0    &W_2-W_3^2&      &1-W_3^2  \nonumber
\end{array}
\right).
\end{eqnarray}
We can calculate the eigenvalues $e_j$ and the numbers $M_j$ 
of the corresponding eigenvectors as 
\begin{eqnarray}
e_1&=&1-W_2=\frac{4m(N-m)}{N(N-1)};\
M_1=N-1,\nonumber\\
e_2&=&1-W_3^2+(N-1)(W_2-W_3^2)=0;\
M_2=1,\nonumber\\
e_3&=&1+W_3+(N-1)W_1=\frac{2N-2m+2mN-2m^2}{N};\
M_3=1,\nonumber\\
e_4&=&1-W_3+(N-1)W_1=\frac{2m+2mN-2m^2}{N};\
M_4=1,\label{lambda4}\\
e_5&=&1+W_3-W_1=\frac{2N^2-2N-4mN+2m+2m^2}{N(N-1)};\
M_5=N-1,\nonumber\\
e_6&=&1-W_3-W_1=\frac{2m^2-2m}{N(N-1)};\
M_6=N-1.\nonumber
\end{eqnarray}
The largest one is $e_3$, 
which degenerates with $e_4$ when $N=2m$.
We thus obtain Eq.~(\ref{e_max}).


\begin{thebibliography}{99}
\bibitem{O(N)}
Throughout this paper, we say that $f(N)=O(g(N))$ if
$\displaystyle \lim_{N\to\infty}f(N)/g(N)={\rm constant}\neq 0$, and that 
$f(N)=o(g(N))$ if $\displaystyle \lim_{N\to\infty}f(N)/g(N)=0$.
\bibitem{molecule}
For example, a ten-atom molecule can be regarded as a single particle, 
ten particles, and much more particles (nuclei and electrons), 
in the energy ranges of $\sim \mu$eV, $\sim$meV, and $\sim$eV, 
respectively.
\bibitem{current}
This statement would be understandable, 
from the discussion in endnote \cite{TD}, 
for macroscopic variables 
that define equilibrium states. 
It is worth mentioning that the statement is also true for most 
macroscopic variables defining non-equilibrium states, such 
as the electric current density $J$.
That is, the macroscopic current density $J$ 
must be an average of the microscopic current density 
$j$ over a macroscopic region.
The spatial average introduces a smoothing effect, 
and $J$ becomes a proper macroscopic variable.
\bibitem{TD}
Some of macroscopic variables, such as the volume and temperature, 
 in thermodynamics cannot be represented as an additive operator.
Although the volume is additive, it is usually considered as a boundary 
condition rather than a quantum-mechanical observable. 
We follow this convention. 
Regarding the temperature, it is a non-mechanical variable
that can be defined only for equilibrium states.
However, we note that at thermal equilibrium
a macroscopic state is, hence the values of 
non-mechanical variables are, uniquely determined
by a set of additive observables and 
the boundary conditions \cite{Tisza,Callen}.
Therefore, 
if two equilibrium states have 
distinct values of a non-mechanical variable 
they must have distinct values of some of the additive observables.
It therefore seems that considering additive observables is sufficient.
\bibitem{beyondscope}
Although other states are possible, they are beyond the scope of 
the present paper.
\bibitem{homogeneous}
For spatially homogeneous states, 
$p=1$ implies that the state has the cluster property.
This is the case in the present paper
because we only study translationally invariant states.
\bibitem{propagate}
Although we have not considered time evolution in this paper, 
the spatial propagation has been incorporated into our analysis for the 
following reason.
A magnon state created by the magnon operator $\hat M_k^\dagger$ 
is an approximate eigenstate of a Hamiltonian of a magnet.
Since an energy eigenstate is a stationary state, 
it is a state `after all possible propagation is finished.'
Therefore, spatial propagation has already been incorporated into 
magnon states.
\bibitem{propagate2}
The generation of entanglement by the Hamilton dynamics was suggested in 
many works, e.g., Ref.~\cite{Leggett1}.
\end{thebibliography}
\end{document}